\documentclass{cimento}

\usepackage{graphicx}  % got figures? uncomment this

\title{The ``Yes, Magellanic Clouds Again'' survey: preliminary results}
\author{M.~Gatto\from{ins:x}\ETC,
V.~Ripepi\from{ins:x},
M.~Tosi\from{ins:y},
M.~Bellazzini\from{ins:y},
M.~Cignoni\from{ins:y}\from{ins:z}\from{ins:w},
C.~Tortora\from{ins:x},
M.~Dall'Ora\from{ins:x}}
\instlist{\inst{ins:x} INAF-Osservatorio Astronomico di Capodimonte, Via Moiariello 16, I-80131, Naples, Italy
\inst{ins:y} INAF-Osservatorio di Astrofisica e Scienza dello Spazio, Via Gobetti 93/3, I-40129 Bologna, Italy
\inst{ins:z} Physics Departement, University of Pisa, Largo Bruno Pontecorvo, 3, I-56127 Pisa, Italy
\inst{ins:w} INFN, Largo B. Pontecorvo 3, I-56127, Pisa, Italy}
\begin{document}

\maketitle

\begin{abstract}
We present preliminary findings from the photometric survey "Yes, Magellanic Clouds Again" (YMCA, PI: V. Ripepi), covering 110 square degrees in the outer regions of the Magellanic Clouds (MCs), a pair of interacting galaxies and the most massive dwarf satellites of the Milky Way. %The survey achieves a notable photometric depth, allowing us to resolve faint, old stellar populations. %The YMCA survey has yielded breakthrough outcomes shedding light on the evolutionary history of the MCs. 
Among the key results, we discovered four star clusters (SCs) within the Large Magellanic Cloud (LMC) exhibiting ages within the so-called ``age gap'', a period deemed so far devoid of SCs. Additionally, we unveiled an ancient stellar system associated with the LMC, featuring structural properties in between the globular clusters and the ultra-faint dwarf galaxies of the Local Group. These discoveries significantly contribute to our understanding of the MCs' evolution and their complex interaction history.
\end{abstract}

\section{Introduction}

The Magellanic Clouds (MCs) stand as a unique laboratory for the study of interacting galaxies, given their proximity ($D \simeq$ 50 kpc for the Large Magellanic Cloud, LMC \cite{degrijs-wicker-bono-2014}, and $D \simeq$ 60 kpc for the Small Magellanic Cloud, SMC \cite{degrijs&bono-2015}). %This close proximity allows us for detailed examinations of merger theories and galaxy evolution.
Additionally, the MCs engage in a dynamic interplay with the Milky Way (MW), %and thus 
representing the closest example of a three-body interacting system. 
Evident signatures of these interactions include the Magellanic Stream, a very extended filamentary structure of mostly HI gas \cite{Donghia&Fox-2016}, and the Magellanic Bridge (MB), %; \cite{Kerr1957}), 
a stream made of gas and stars linking the SMC East side with the LMC West side.
In recent years, numerous studies have sought to reconstruct the evolutionary history of the MCs, by exploiting their star formation history (SFH; e.g. \cite{Harris2009, Cignoni2013}) or the age distribution of their star clusters (SCs; e.g. \cite{Nayak2016, Pieres2016}).
Despite significant strides in elucidating the MCs' recent past, particularly supporting a scenario where the LMC and SMC became an interacting pair only a few Gyr ago \cite{Diaz&Bekki2012}, several critical questions persist.
Notably, the almost total absence of SCs in the so-called ``age gap'' of the LMC, an age interval ranging from $\sim 4$ to $\sim 10$ Gyr \cite{Jensen1988}, remains unexplained.
As the LMC star field population exhibits no such gap \cite{Tosi2004, Piatti&Geisler2013}, the SC formation, and the SFH of the LMC seem to be decoupled, making it peculiar when compared to other galaxies of the Local Group.
Within this context, a complete census of SCs with accurate age estimates is essential to unravel the full evolutionary history of the LMC.
While recent deep surveys with high spatial resolution have increased the total number of LMC SCs, the majority of the studies have focused on the LMC main body, leaving the area beyond $\simeq$ 4° from the center largely unexplored \cite{Nayak2016}.
Here we present preliminary results from the research of undetected SCs %and the analysis of known SCs 
within the footprint of the ``Yes, Magellanic Clouds Again'' survey (YMCA; PI: V. Ripepi). By extending the scrutiny to the outskirts of the LMC we enhance our knowledge of the LMC SC system and contribute valuable insights into the intricate interaction history of the MCs.

\section{The YMCA survey}

The optical photometric YMCA survey is a project undertaken during the Guaranteed Time Observation (GTO) allocation and executed utilizing the VLT survey (VST) telescope \cite{Capaccioli&Schipani2011}.
This survey probes the peripheries of both the SMC and the LMC, including a strip above the MB that connects them (see Fig.~1 in \cite{Gatto-2020}).
YMCA covers 110 square degrees in the $g$ and $i$ filters, achieving a photometric depth of approximately 1.5-2 magnitudes ($g \simeq$ 24 mag) below the main sequence turn-off (MSTO) of the oldest stellar population of the MCs, enabling the exploration of even the earliest phases of the MCs.
Its primary objective is to unveil the evolutionary history of the MCs, while also refining our temporal understanding of their past interactions.
Notably, YMCA investigates regions that have not yet been observed with deep and homogeneous photometry. %, rendering it the inaugural catalog of its kind in the outskirts of the LMC and SMC.
%Observations commenced during Period 98 and concluded in Period 108 with the observation of the final tile.

\section{Detection and analysis of SCs}

To discover new SCs we developed an algorithm that pinpoints regions of the sky where the stellar density surpasses the local background, operating with the star coordinates as its sole input parameter (see \cite{Gatto-2020} for a comprehensive description of the algorithm).
%In this section, we present a compact overview of the algorithm which is comprehensively detailed in .
%The algorithm operates , independently processing each tile to  
To accomplish this, it employs a Kernel Density Estimation technique, utilizing both the \emph{tophat} and \emph{gaussian} functions as kernels, with a fixed bandwidth of 0.2'.
%Subsequently, the algorithm automatically computes the centers and radii of the identified over-densities, as elaborated in Sect.~3.2 of \cite{Gatto-2020}.
For determining the age of each SC candidate, we employed a visual isochrone matching procedure. PARSEC isochrones \cite{Bressan-2012} were utilized, with the distance modulus fixed at ${\rm DM} = 18.49$ mag (i.e. $\simeq$ 50 kpc, \cite{degrijs-wicker-bono-2014}). The metal content was set based on the age-metallicity relation derived by \cite{Piatti&Geisler2013}. The age of the isochrones was varied systematically to identify the best match with the positions of stars on the de-reddened Color-Magnitude Diagram (CMD). 
%To account for uncertainties, 
For each SC, we estimated an age uncertainty of $\sigma_{\log t} = 0.2$ dex, combining both statistical errors from the visual fitting and systematic errors due to the actual extension of the LMC along the line of sight.

\section{The age distribution of the SCs in the periphery of the LMC}

\begin{figure}
    \centering
    \includegraphics[width=0.9\textwidth]{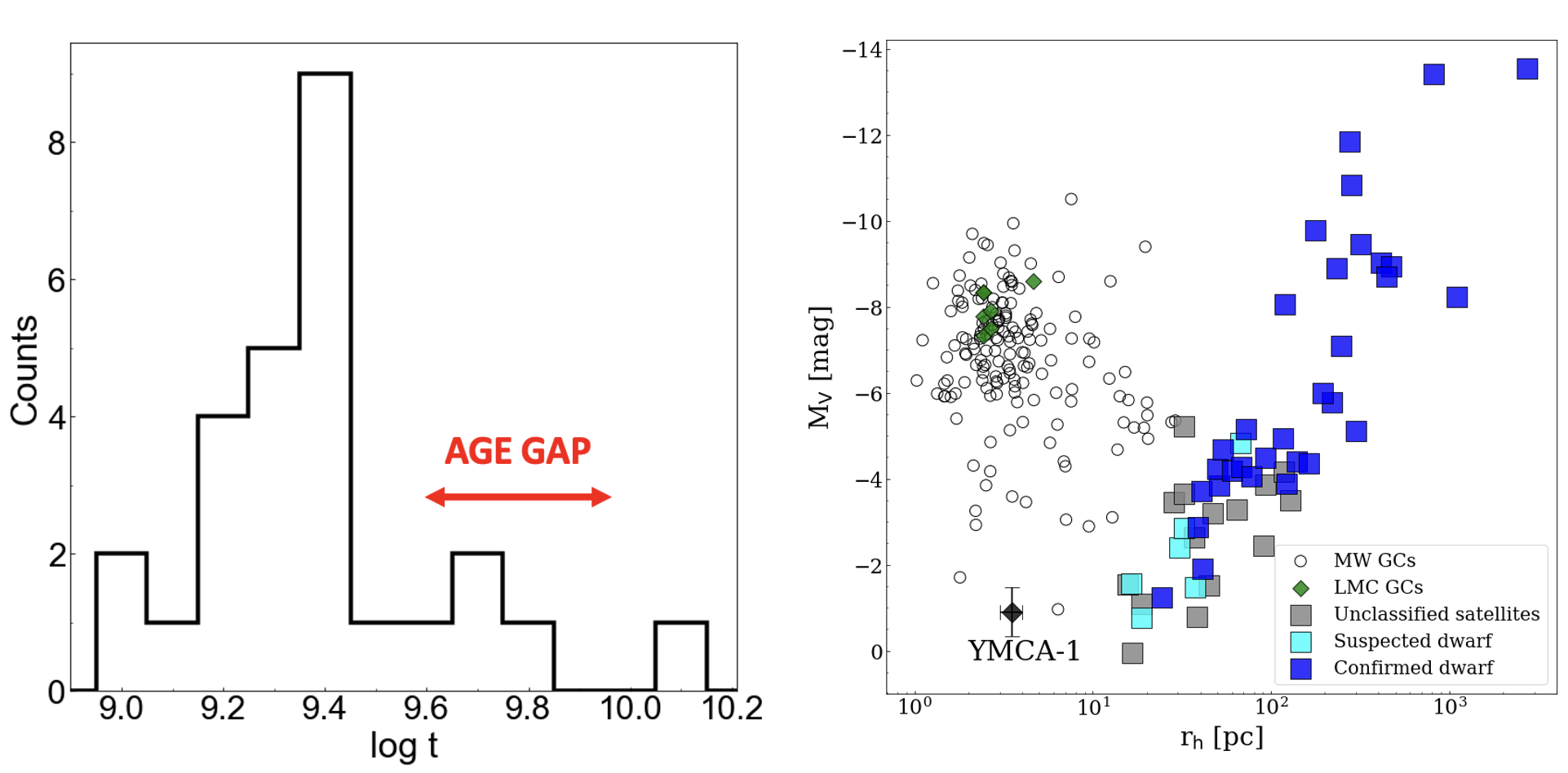}
    \caption{\emph{Left:} Age distribution of 31 SCs located in the LMC periphery. \emph{Right:} $M_V$ vs $r_h$ for YMCA-1 (black diamond), some LMC GCs (green diamonds; values taken from \cite{Piatti&Mackey2018, Mackey&Gilmore2003}), MW GCs (empty circles; values taken from \cite{Baumgardt&Hilker2018,Koposov-2007}), and UFDs (colored squares) reported in \cite{Simon2019}.}
    %\includegraphics[width=0.45\textwidth]{mag_vs_rh_cimento.png}
    %\caption{Age distribution of the 31 SCs located in the LMC periphery.}
    
    \label{fig:sc_age}
\end{figure}

The left panel of Figure~\ref{fig:sc_age} displays the age distribution of the 31 candidate SCs detected in the outskirts of the LMC. %, spanning across 79 YMCA tiles. 
Notably, 12 of these candidates have never been previously reported in the literature, thereby increasing the count of known SCs in these fields by $\simeq$ 60\%.
The distribution prominently features a peak at $\simeq$ 2.5 Gyr, consistent with the observations reported in \cite{Pieres2016, Gatto-2020}. %Moving towards younger ages ($t \leq 2$ Gyr), the number of SCs gradually decreases, and conversely, after $\simeq$ 3 Gyr, just before the onset of the age gap, it sharply declines.
This notable spike in SC formation potentially indicates a past close fly-by between the MCs, serving as a trigger for a new episode of massive SC formation. This interpretation aligns with various studies reconstructing the orbital history of the MCs, suggesting a pericentric passage between the MCs between 1 and 3 Gyr ago \cite{Patel-2020}.
Even more importantly, Figure~\ref{fig:sc_age} reveals the presence of four candidate SCs within the age gap, three of which are new discoveries. %, particularly those with ages ranging from 5.5 to 6.0 Gyr.
The fourth SC is KMHK~1762 that we re-analyzed with YMCA photometry, demonstrating that it is older than previously thought, being the third ever confirmed age gap SC \cite{Gatto-2022a}.
In light of these findings, % and the comprehensive discussion in \cite{Gatto-2020}, we 
we posit that the age gap %in the SC population of the LMC 
may be narrower than previously thought (between 7 - 10 Gyr), or it could potentially be an observational bias arising from the limitations of previous surveys in terms of photometric depth and spatial coverage.
Confirmation of these results would close a controversial topic about a different evolutionary path between SCs and stellar field components in the LMC, which lasts since more than 30 years.

\section{YMCA - 1: a peculiar SC of the LMC}

The SC finder algorithm applied on the YMCA tiles spotted also a very peculiar stellar system, that we dubbed YMCA-1 \cite{Gatto-2022b}. 
The VST data only revealed a few stars on the red-giant branch and the top main sequence (MS), and were not sufficiently deep %to unambiguously establish its real physical nature, lacking the depth necessary 
to unequivocally %confirm it as a stellar system and to 
determine a robust distance measurement.
To address this, we conducted a deep photometric follow-up utilizing the FORS2@VLT instrument.
The resulting CMD of YMCA-1 %at the light of the deeper VLT data 
displays a well-defined MS extending down to $g \simeq$ 26 mag (see Fig.~2 in \cite{Gatto-2022b}), roughly 2.5 mag below the MSTO. %Leveraging the Automated Stellar Cluster Analysis package (ASteCA, \cite{Perren-2015}) on the CMD, 
We conducted an automated search for the best-fitting isochrone model with the Automated Stellar Cluster Analysis package (ASteCA, \cite{Perren-2015}).
The fit yielded a distance modulus of $\mu_0 = 18.72^{+0.15}_{-0.17}$~mag, corresponding to $\sim$ 55 kpc, placing YMCA-1 behind the LMC main disc. Additionally, the age of YMCA-1 was estimated to be $t \sim 11.7^{+1.7}_{-1.3}$ Gyr, with a metallicity of [Fe/H] $\simeq -1.12^{+0.21}_{-0.13}$ dex.
%Further exploration of YMCA-1's properties involved a comparison with the 15 known old globular clusters (GCs) of the LMC. 
The right panel of Figure~\ref{fig:sc_age} compares the total luminosity in the $V$ band and half-light radius of YMCA-1 with those of some old LMC GCs, MW GCs and ultra-faint dwarf galaxies (UFDs). The stark contrast between YMCA-1 and LMC GCs, which inhabit a different region of the $M_V$-$r_h$ plane, suggests that YMCA-1 may belong to a distinct sub-class of stellar systems within the LMC, sharing features with both classical GCs and UFDs.
%To summarize, YMCA-1 likely represents an old LMC GC with peculiar characteristics. 
%While spectroscopic follow-up could confirm its association with the LMC, until such confirmation is obtained, the less likely hypothesis that YMCA-1 is a remote MW GC cannot be entirely ruled out.

\section{Summary}

We presented preliminary results from the optical photometric survey ``Yes, Magellanic Clouds Again'', conducted with the VST, targeting 110 square degrees in the $g$ and $i$ filters within the outskirts of both the LMC and the SMC. 
Specifically, we presented the outcomes of an extensive exploration in search of new SCs in the outskirst of the LMC, achieved by means of a fine-tuned cluster finder algorithm.
The catalog of newly identified SCs in the outer regions of the LMC challenges the presumed existence of the age gap, proposed over three decades ago. This finding holds the promise for reconciling observed discrepancies between the star field and SC formation histories over this extended period.
Additionally, the SC age distribution showcases a pronounced peak at $\simeq$ 2.5 Gyr, hinting for a past close encounter between the MCs that triggered a new intense SC formation activity.
Our search also led to the discovery of YMCA-1, an ancient, faint, and compact SC likely associated with the LMC. YMCA-1 is a very intriguing stellar system as it displays structural properties intermediate between those of the faintest GCs and UFDs.
A crucial next step involves spectroscopic follow-up observations to definitively determine the nature of YMCA-1.
A forthcoming paper will present the comprehensive final data release of the YMCA survey, including the analysis of the stellar field component in the periphery of the MCs.

\acknowledgments
This work is based on DDT allocated by ESO (program 108.23LX.001) and on INAF-VST GTO programmes. M.G. acknowledges the ``IAF AstroFIt'' grant (1.05.11).

\end{document}